\documentclass[preprint, 12pt]{elsarticle}
\biboptions{sort&compress}

\usepackage{amsmath}
 
\newcommand{\Sb}{La$_{0.5}$Rh$_4$Sb$_{12}$}
\newcommand{\RhP}{La$_{0.6}$Rh$_4$P$_{12}$}
\newcommand{\As}{La$_{0.5}$Rh$_4$As$_{12}$}

\newcommand{\Sbx}{La$_x$Rh$_4$Sb$_{12}$}

\newcommand{\Asx}{La$_x$Rh$_4$As$_{12}$}

\newcommand{\fs}{$R_xM_4X_{12}$}
\newcommand{\kb}{\ensuremath{k_{\rm B}}}

\newcommand{\tc}{\ensuremath{T_{\it c}}}
\newcommand{\eg}{\ensuremath{E_{\rm g}}}

\journal{J. Alloys Compd.}

\begin{document}

\begin{frontmatter}

\title{Electrical and magnetic properties of {\Sb} filled skutterudite synthesized at high pressure}

\author{S.~Ibuka\corref{cor1}\fnref{pakek}}
\ead{ibuka@post.j-parc.jp}
\author{M.~Imai}
\author{M.~Miyakawa}
\author{T.~Taniguchi}
\address{National Institute for Materials Science, Tsukuba, Ibaraki 305-0047, Japan}
\cortext[cor1]{Corresponding author}
\fntext[pakek]{Present address: High Energy Accelerator Research Organization, Tokai, Ibaraki 319-1106, Japan.}

\begin{abstract}
A filled skutterudite, {\Sb}, with a lattice constant of 9.284(2) {\AA} was synthesized using a high-pressure technique. The electrical resistivity showed semiconducting behavior and the energy gap was estimated to be more than 0.08 eV. Magnetic susceptibility measurements indicated temperature-independent diamagnetism, which originates from Larmor diamagnetism. The electrical properties of this compound are more similar to those of the {\As} semiconductor with an energy gap of 0.03 eV than to those of the {\RhP} superconductor.
\end{abstract}


\end{frontmatter}

\section{Introduction}
Filled skutterudites {\fs} ($R =$ rare-earth; $M =$ Fe, Ru, Os, etc.; $X =$ P, As, and Sb), which crystallize with the LaFe$_4$P$_{12}$-type structure (Space group: $Im\bar{3}$, $Z = 2$)~\cite{Jeit77}, exhibit a variety of physical properties, including superconductivity, semiconductivity, ferromagnetism and antiferromagnetism, depending on the combination of $R$, $M$ and $X$~\cite{Sale03,Sato09}. 

The highest superconducting transition temperature {\tc} among {\fs} was 10.3~K for LaRu$_4$As$_{12}$~\cite{Shir97} before the discovery of superconductivity with $\tc = 17$~K for {\RhP}, which was synthesized using the high-pressure technique reported by Shirotani and colleagues in 2005~\cite{Shir05, Take07, Imai07, Imai07b}. The relatively high {\tc} in the latter filled skutterudite has encouraged materials scientists to search for new superconductors with similar compositions. 
One feature of {\RhP} is that it includes a cobalt group element, $M =$ Rh; most of the existing filled skutterudites involve iron group elements.
If filled skutterudites with cobalt group elements are synthesized under ambient pressure, then the site occupancy of the $R$ atoms will become rather low, as observed for La$_{0.2}$Co$_4$P$_{12}$~\cite{Zemn86} and La$_{0.05}$Rh$_4$Sb$_{12}$~\cite{Zeng00}. 
Thus, the high site occupancy of {\RhP} can be attributed to high-pressure synthesis, and is believed to be essential for superconductivity. 
A new candidate superconductor is {\Sbx}, because {\As} synthesized at high pressure was reported not to be a superconductor, but rather a semiconductor with a narrow energy gap {\eg} of 0.03~eV~\cite{Arii09,Arii10}.
Therefore, with the aim of finding a new superconductor, {\Sbx} filled skutterudites were synthesized utilizing high-pressure techniques. 
The synthesis conditions were optimized to reduce the impurity phases and obtain a sample with a high La-site occupancy.
Electrical resistivity and magnetic susceptibility measurements were then performed to verify whether {\Sbx} is a superconductor. 

\section{Materials and methods}
To determine the optimal conditions for the solid-state synthesis of polycrystalline {\Sbx}, samples were synthesized using starting materials with various atomic ratios at different temperatures and pressures. The purity of the starting materials was 99.9\% for La, 99.9\% for Rh and 99.999\% for Sb.
Rh$_{3.7}$Sb$_{12}$, which consists of RhSb$_3$ and Sb, was prepared in advance by solid-state reaction at ambient pressure. These materials were weighed with molar ratios of La:Rh:Sb = 1:4:12 and 0.5:3.7:12, La:Rh$_{3.7}$Sb$_{12}$ = 0.5:1, and La:Sb = 1:3 in an argon-gas-filled glove box, and ground for 3 min using a vibrating mill. The following procedures were different between the high-pressure and ambient pressure syntheses. For high-pressure synthesis, the resulting powder was pressed into a pellet, placed in a hexagonal boron nitride capsule, sealed in a gold capsule under an argon gas atmosphere, and heat-treated at 1073 or 1173~K under pressures of 5.5 or 7.7~GPa for 2~h using a belt-type high-pressure apparatus~\cite{Yama92}. For ambient pressure synthesis, a pellet of the milled materials was placed in an alumina crucible, sealed in a quartz tube under an argon gas atmosphere, and heat-treated at 1073~K for 100~h in an electric furnace. The samples were characterized using powder X-ray diffraction (XRD; RINT TTR-III, Rigaku) with Cu K$\alpha$ radiation (40~kV/150~mA). A one-dimensional position-sensitive Si detector was used. The chemical composition was determined from wavelength-dispersive X-ray spectroscopy (WDS) measurements using an electron probe microanalyzer (JXA-8500F, Jeol) and by energy-dispersive X-ray spectroscopy (EDS) with a scanning electron microscope (SU70, Hitachi High-Technologies). For physical property measurements, the sample with the least impurity phases was selected. DC magnetization measurements were conducted using a commercial superconducting quantum interference device (SQUID) magnetometer (MPMS, Quantum Design) under an applied field of $H = 10000$~Oe from room temperature down to 2.0~K. Four-probe electrical resistivity measurements were performed using an in-house-built apparatus with a DC current density of 0.1~mA/mm$^2$. A $^4$He closed-cycle cryostat, of which the lowest temperature was 2.8~K, was used to cool the sample.

\section{Results}
\subsection{Optimization of synthesis conditions}
\begin{figure}
	\begin{center}
    \includegraphics[bb=0 0 288 403, width=0.8\hsize]{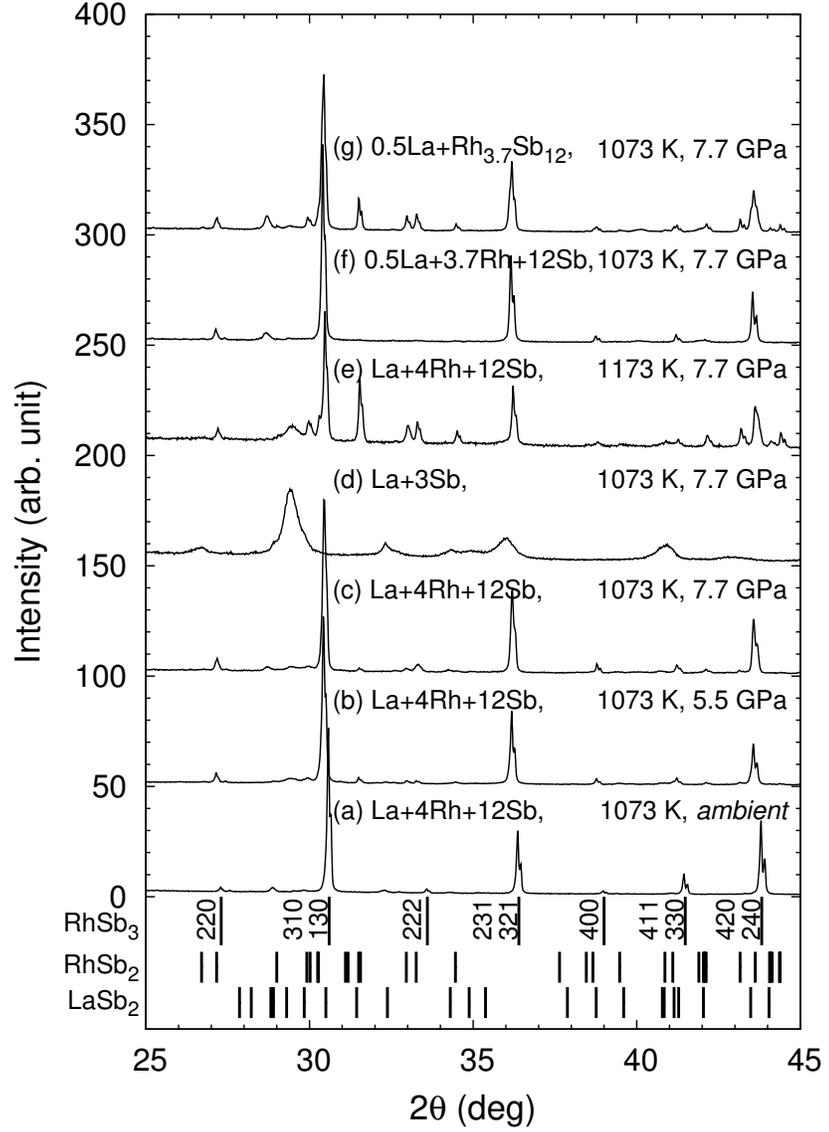}
    \caption{\label{fig1} XRD patterns for the prepared samples. For samples (a)-(c), the starting materials and synthesis temperature were fixed at La:Rh:Sb = 1:4:12 and 1073~K, while the synthesis pressures were ambient pressure, 5.5 and 7.7~GPa, respectively. For sample (d), the synthesis conditions were La:Sb = 1:3, 1073~K and 7.7~GPa. For sample (e), the synthesis conditions were La:Rh:Sb = 1:4:12, 1173~K and 7.7~GPa. For samples (f)-(g), the synthesis temperature and pressure were fixed at 1073~K and 7.7~GPa, while the starting materials were La:Rh:Sb = 0.5:3.7:12 and La:Rh$_{3.7}$Sb$_{12}$ = 0.5:1, respectively. The solid bars represent the Bragg peak positions for RhSb$_3$, RhSb$_2$ and LaSb$_2$. For RhSb$_3$, the index numbers are noted beside the bars.} 
    \end{center}
\end{figure}

\begin{figure}
	\begin{center}
        \includegraphics[bb=0 0 1202 1073, width=0.9\hsize]{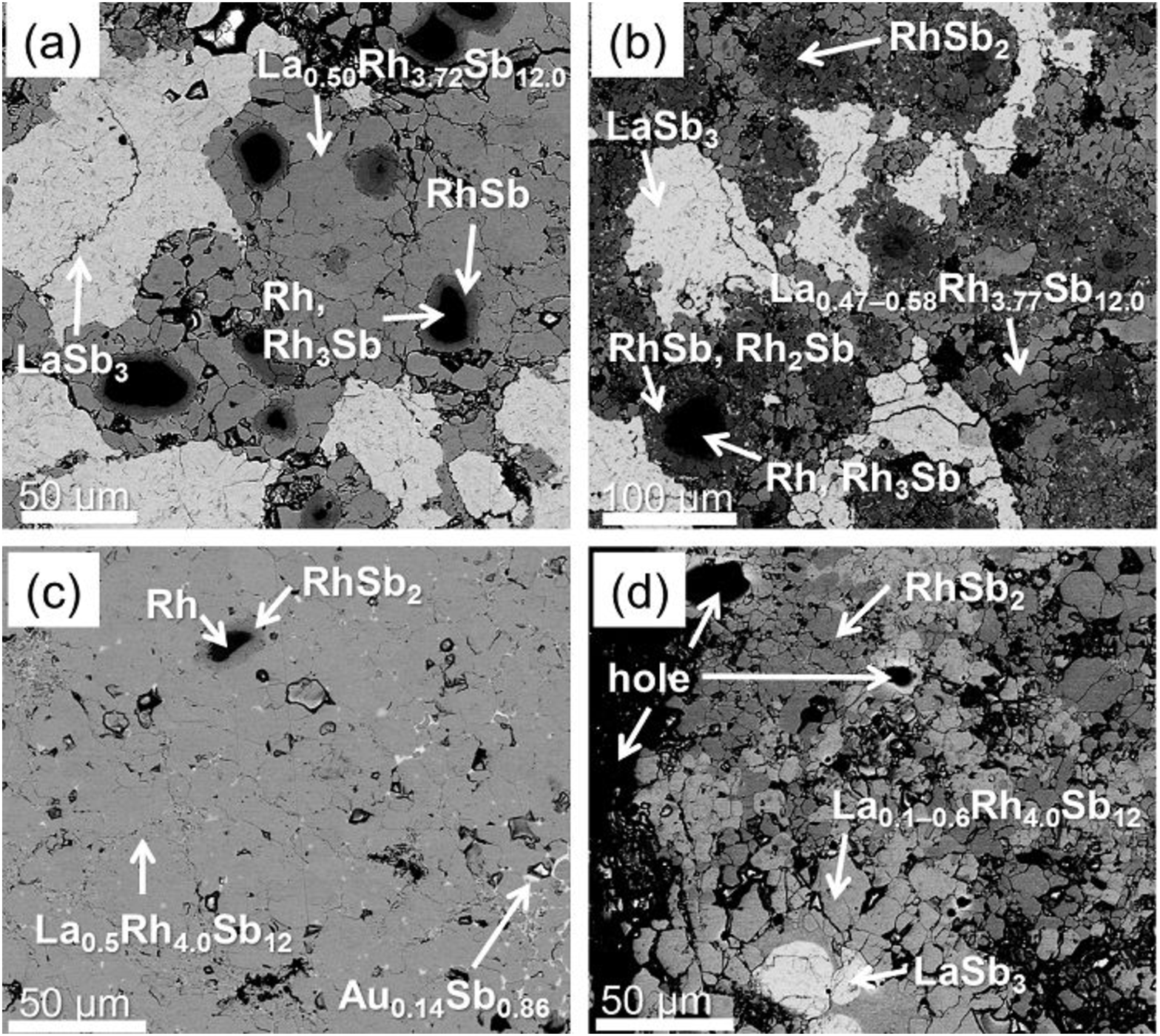}
        \caption{\label{fig2} BEC images of samples prepared at 7.7~GPa with a starting material ratio of La:Rh:Sb = 1:4:12 at (a) 1073~K and (b) 1173~K, and at 7.7~GPa and 1073~K for (c) La:Rh:Sb = 0.5:3.7:12 and (d) La:Rh$_{3.7}$Sb$_{12}$ = 0.5:1.} 
	\end{center}
\end{figure}
The synthesis pressure was initially optimized for a synthesis temperature of 1073 K with a stoichiometric 1:4:12 molar mixture of La, Rh and P as starting materials. Figs.~\ref{fig1}(a), (b) and (c) show XRD patterns of the three samples synthesized at ambient pressure, and at 5.5 and 7.7~GPa, respectively. All the strong peaks are indexed to RhSb$_3$~\cite{Kjek74}, which indicates that RhSb$_3$ or {\Sbx} was formed. It should be noted that filled skutterudites have similar diffraction patterns to unfilled skutterudites because they both have the same space group of $Im\bar{3}$~\cite{Take07}. The $a$ lattice constants were estimated to be 9.231(1), 9.285(1) and 9.285(1) {\AA} for the samples synthesized at ambient pressure, 5.5 and 7.7~GPa, respectively. The $a$ lattice constant for the sample synthesized at ambient pressure is the same as that for RhSb$_3$(9.232~{\AA})~\cite{Kjek74}, while those for the samples synthesized at 5.5 and 7.7~GPa are significantly larger than that for RhSb$_3$. 
The weak extra XRD peaks for the samples produced at ambient pressure and at 5.5 and 7.7~GPa are attributed to LaSb$_2$ and RhSb$_2$, respectively. 
There was little difference between the samples synthesized at 5.5 and 7.7~GPa; therefore, 7.7~GPa was selected as the synthesis pressure for further experiments.
A backscattered electron composition (BEC) image of the sample prepared at 7.7~GPa is shown in Fig.~\ref{fig2}(a). The composition of the gray area was determined by WDS to be La$_{0.50}$Rh$_{3.72}$Sb$_{12.0}$.
It should be noted that the compositional formula for the filled skutterudite is described with the molar ratio of Sb being 12 in this article. 
Both the composition and XRD spectrum indicate that the filled skutterudite was successfully synthesized.
The La site occupancy $x = 0.50$ was higher than that for {\Sbx} ($x = 0.05$) synthesized at ambient pressure~\cite{Zeng00}, which indicates that high pressure is effective for La insertion into unfilled skutterudites.
In addition, these results suggest that the $a$ lattice constant increases with $x$.
A large amount of impurity phases was also observed in Fig.~\ref{fig2}(a). The white, dark gray and black areas were assigned to LaSb$_3$, RhSb and a mixture of Rh and Rh$_3$Sb, respectively. The XRD pattern of the largest impurity phase, LaSb$_3$, has not been reported; therefore LaSb$_3$ was synthesized at 7.7~GPa and 1073~K with the starting material ratio La:Sb = 1:3, and was confirmed by BEC imaging to consist of single phase LaSb$_3$ (not shown). The XRD pattern for LaSb$_3$ is presented in Fig.~\ref{fig1}(d). The peaks are broad, which indicates that the structural correlation length is short. These characteristics made it difficult to detect LaSb$_3$ impurity using XRD for the {\Sbx} sample (Fig.~\ref{fig1}(c)), although a large amount of LaSb$_3$ was present.

The synthesis temperature was subsequently optimized to 7.7~GPa. Figs.~{\ref{fig1}(e) and {\ref{fig2}}(b) present an XRD pattern and a BEC image, respectively, for the sample synthesized at 1173~K using a starting material ratio of La:Rh:Sb = 1:4:12. Besides peaks for {\Sbx}, strong RhSb$_2$ peaks are observed in Fig.~{\ref{fig1}}(e). La$_{0.47-0.58}$Rh$_{3.77}$Sb$_{12.0}$, for which the composition was determined by WDS, occupies the small area in Fig.~\ref{fig2}(b), and impurities of LaSb$_3$, Rh, and Rh-Sb phases such as RhSb$_2$, RhSb, Rh$_2$Sb, Rh$_3$Sb occupy a large area. The volume fraction of impurities is larger for the sample prepared at the higher temperature of 1173~K than that at 1073~K, which indicates incongruent melting of {\Sbx}. For these reasons, 1073~K was selected as the optimal synthesis temperature. The appearance of a large amount of LaSb$_3$ impurity phase in both samples was attributed to an excess supply of La. Rh and the surrounding Rh-Sb phase were deduced to be non-equilibrium phases that appear due to the high melting temperature of Rh and the finite reaction time. 

Finally, the starting material ratio was optimized by fixing the synthesis pressure and temperature at 7.7~GPa and 1073~K. Two starting material mixtures were attempted: a 0.5:3.7:12 molar mixture of La, Rh, and Sb and a 0.5:1 molar mixture of La and Rh$_{3.7}$Sb$_{12}$. Figs.~{\ref{fig1}}(f) and (g) show XRD patterns for the samples synthesized from these mixtures at 7.7~GPa and 1073~K. The peak intensity for the impurity phases is much smaller in the XRD pattern for the 0.5:3.7:12 molar ratio sample than that for the 1:4:12 molar ratio. BEC images of the sample synthesized using La:Rh:Sb = 0.5:3.7:12 and that using La:Rh$_{3.7}$Sb$_{12}$ = 0.5:1 are presented in Figs.~\ref{fig2}(c) and (d), respectively. The chemical compositions were determined from EDS measurements. The sample synthesized with the 0.5:3.7:12 molar ratio for La, Rh, and Sb consists of almost single phase La$_{0.5}$Rh$_{4.0}$Sb$_{12}$, although a small amount of the Au$_{0.14}$Sb$_{0.86}$ impurity phase is evident~\cite{Gies68, Iyo14}, which could be due to contamination from the gold cell. In the sample started from Rh$_{3.7}$Sb$_{12}$, a large amount of impurity phase was observed as shown in Fig.~{\ref{fig2}}(d), which is consistent with the XRD pattern in Fig.~{\ref{fig1}}(g). The La content $x$ in {\Sbx} fluctuates between 0.1 and 0.6, which indicates that the phase has not reached the equilibrium state within the heating time of 2~h. This may be attributed to the slow reaction between La and RhSb$_3$ due to the high stability of RhSb$_3$ compared with Rh and Sb.} As a result, the appropriate conditions for the synthesis of {\Sbx} were determined to be 1073~K and 7.7~GPa with a starting material ratio of La:Rh:Sb = 0.5:3.7:12. 

  The sample synthesized under the optimized conditions had a lattice constant of 9.284(2) {\AA} and the chemical composition was 3.9(3) wt\% La, 21.1(2) wt\% Rh, and 74.9(2) wt\% Sb, which corresponds to the chemical formula La$_{0.55(5)}$Rh$_{3.99(4)}$Sb$_{12.0(2)}$.

\subsection{Physical property measurements}
\begin{figure}
\begin{center}
        \includegraphics[bb= 0 0 288 360, width=0.8\hsize]{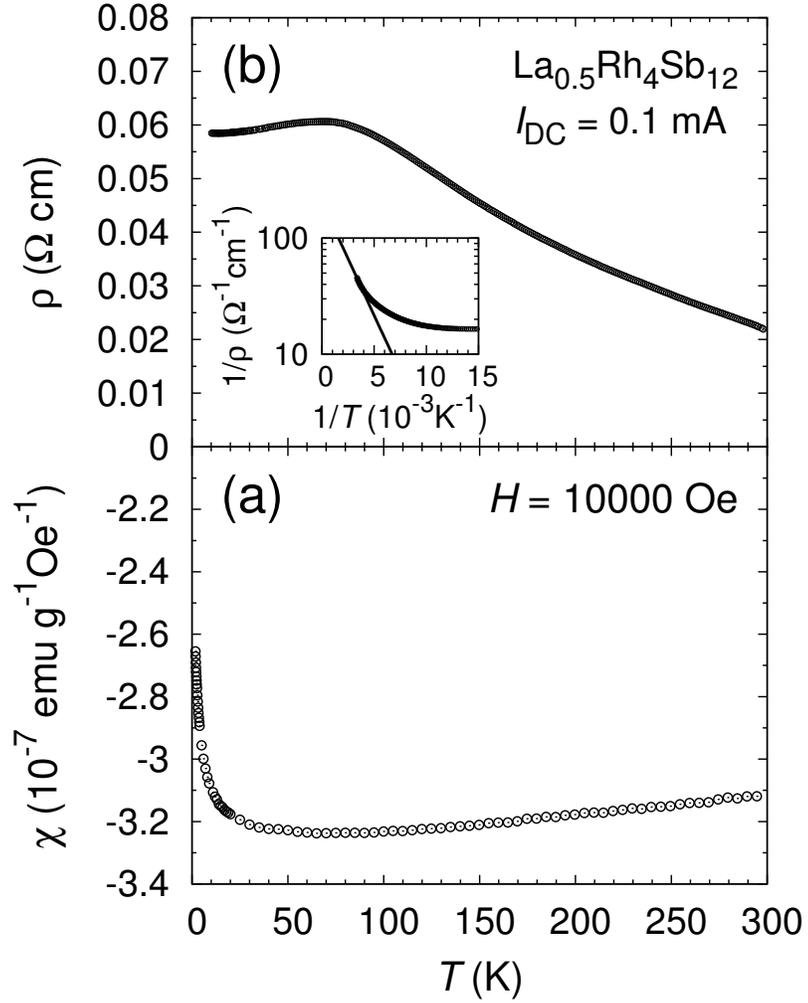}
        \caption{\label{physProp} (a) Temperature $T$ dependence of the magnetic susceptibility ${\chi}$ of {\Sb} from 2~K to room temperature. (b) $T$ dependence of the electrical resistivity $\rho$ for {\Sb} from 2.8~K to room temperature. The inset shows the inverse temperature dependence of the conductivity, $1/\rho$, where the solid line represents a fit with the exponential function above 270~K.}
\end{center}
\end{figure}
The sample prepared under the optimized conditions was used for physical property measurements. Fig.~{\ref{physProp}}(a) shows the temperature $T$ dependence of the magnetic susceptibility $\chi$ for {\Sb} from 2~K to room temperature, which indicates weak diamagnetism that is independent of $T$. The absence of perfect diamagnetism indicates that {\Sb} is not a superconductor. In addition, the lack of paramagnetism derived from free electrons implies that this is not a metal, but a semiconductor or insulator. The diamagnetism was approximately $-3.1{\times}10^{-7}$~emu/(gOe), which corresponds to $-6{\times} 10^2$~cm$^3$/mol for {\Sb}. This value is comparable to Larmor diamagnetism. The weak increase at low temperatures may be due to magnetic impurities. 

Fig.~{\ref{physProp}}(b) shows the $T$ dependence of the electrical resistivity $\rho$ for {\Sb}. 
At $T > 80$~K, $\rho$ decreases with increasing temperature, showing semiconducting behavior, which is consistent with the diamagnetic susceptibility. 
In addition, the $\rho(T)$ curve has a plateau below 80~K, which may originate from extrinsic semiconductivity, with the carriers being supplied by impurities acting as dopants. This type of $\rho-T$ dependence has been observed in narrow-gap semiconductors such as SrSi$_2$~\cite{Imai07c} and FeSi~\cite{Pasc97}. 
The conductivity $1/\rho$ could not be fitted above 100~K using either the exponential function $\exp(-{\eg}/{\kb}T)$ or the variable range hopping model function $\exp(-(T_{\rm 0}/T)^{1/(1+d)})$~\cite{Mott68}, where {\kb} is the Boltzmann constant, $T_{\rm 0}$ is a characteristic temperature, and $d$ is the dimension ($d = 1, 2$ and 3). Fitting with the exponential function above 270~K, as shown in the inset of Fig.~{\ref{physProp}}(b), gives an energy gap of ${\eg} = 0.08$~eV, which indicates that {\eg} is greater than 0.08~eV. 

\section{Discussion}
The La site occupancy $x$ is approximately the same for La$_x$Rh$_4X_{12}$ synthesized at high pressures: {\RhP}~\cite{Shir05}, {\As}~\cite{Arii09}, and {\Sb}. These are classified into two groups according to their electronic properties; {\RhP} is a superconductor, while {\As} and {\Sb} are semiconductors with ${\eg} = 0.03$~eV~\cite{Arii10} and ${\eg} > 0.08$~eV, respectively.
The corresponding unfilled skutterudites are similarly classified; an unfilled skutterudite RhP$_3$, which corresponds to {\RhP}, has been experimentally determined to be a metal~\cite{Odil78,Saet09}. RhAs$_3$ and RhSb$_3$, which correspond to {\As} and {\Sb}, respectively, are semiconductors with ${\eg} > 0.85$~eV~\cite{Cail95} and ${\eg} = 0.8$~eV~\cite{Nola96,Cail96}, respectively.
Thus, {\As} and {\Sb} not only have similar electrical properties, but also exhibit similar changes in the electrical properties as a result of La filling. This suggests that La insertion into unfilled metallic skutterudites will be a future strategy for producing new superconductors.

The three compounds LaRu$_4X_{12}$ ($X =$ P, As, Sb) are all superconductors~\cite{Meis81,Shir97,Uchi99}, which implies that the superconductivity is insensitive to the difference of the pnictogen species with high La-site-occupancy. 
Thus, if higher occupancy is realized by higher pressures, superconductivity may be induced in {\Sbx} and {\Asx}, as in {\RhP}. On the assumption that $x$ increases in proportion to the pressure, full occupancy with x = 1 in {\Sbx} would be realized at approximately 15~GPa, which is beyond the achievable pressure of the apparatus used in the present study.

\section{Summary}
A {\Sb} filled skutterudite with a high La concentration was successfully synthesized by utilizing a high-pressure technique. This adds a new example to the filled skutterudites that include cobalt group elements. In addition, the presence of the LaSb$_3$ binary phase at high pressure was confirmed. Electrical resistivity and magnetic susceptibility measurements revealed that {\Sb} is not a superconductor, but a semiconductor with an energy gap greater than 0.08~eV. The electrical properties are more similar to those of {\As} than those of {\RhP}.

\section{Acknowledgments}
The authors thank M. Nishio for experimental support and H. Kitaguchi for allowing use of the SQUID magnetometer. This work was supported by the Funding Program for World-Leading Innovative R\&D on Science and Technology (FIRST), Japan.

\bibliographystyle{model1-num-names}
\bibliography{R1504457-1A}
\end{document}